\documentclass[9pt,shortpaper,twoside,web]{IEEEtran}
\usepackage{cite}
\usepackage{amsmath,amssymb,amsfonts}
\usepackage{algorithmic}
\usepackage{graphicx}
\usepackage{textcomp}
\usepackage{subfig}
\def\BibTeX{{\rm B\kern-.05em{\sc i\kern-.025em b}\kern-.08em
    T\kern-.1667em\lower.7ex\hbox{E}\kern-.125emX}}
\begin{document}
\title{A deep network for sinogram and CT image reconstruction}
\author{Wei Wang, Xiang-Gen Xia, Chuanjiang He, Zemin Ren, Jian Lu,  Tianfu Wang and  Baiying Lei
\thanks{This work was supported by the China Postdoctoral Science Foundation
	(2018M64081) and National Natural Science Foundation of China (No. 61871274). }
\thanks{Wei Wang, Baiying Lei, and Tianfu Wang  are with the School of Biomedical Engineering, Shenzhen University,
	National-Regional Key Technology Engineering Laboratory for Medical Ultrasound,Guangdong Key Laboratory for Biomedical Measurements and Ultrasound Imaging,School of Biomedical Engineering, Health Science Center, Shenzhen University, Shenzhen, China. (e-mail: wangwei@szu.edu.cn, leiby@szu.edu.cn, tfwang@szu.edu.cn). }
\thanks{Xiang-Gen Xia is with the College of Information Engineering, Shenzhen University, Shenzhen, China. He is also with College of Electrical and Computer Engineering, University of Delaware, Newark, DE 19716, USA.  (e-mail: xxia@ee.udel.edu).}
\thanks{Chuanjiang He is with the
College of Mathematics and Statistics,Chongqing University,Chongqing, China (e-mail: cjhe@cqu.edu.cn).}
\thanks{Zemin Ren is with the
	College of Mathematics and Physics,Chongqing University of Science and Technology,Chongqing, China (e-mail: zeminren@cqu.edu.cn).}
\thanks{Jian Lu is with the
	Shenzhen Key Laboratory of Advanced Machine Learning and Applications, Shenzhen University, Shenzhen, China (e-mail: jianlu@cqu.edu.cn).}
}

\maketitle

\begin{abstract}
A CT image can be well reconstructed when the sampling rate of the sinogram satisfies the Nyquist criteria and the sampled signal is noise-free. However, in practice, the sinogram is usually contaminated by noise, which degrades the quality of a reconstructed CT image. In this paper, we design a deep network for sinogram and CT image reconstruction. The network consists of two cascaded blocks that are linked by a filter backprojection (FBP) layer, where the former block is responsible for denoising and completing the sinograms while the latter is used to removing the noise and artifacts of the CT images. Experimental results show that the reconstructed CT images by our methods have the highest PSNR and SSIM in average compared to  state of the art methods.
\end{abstract}

\begin{IEEEkeywords}
Low-dose CT, deep learning, auto-encoder,
convolutional, deconvolutional, residual neural network
\end{IEEEkeywords}

\section{Introduction}
\label{sec:introduction}
X-ray computed tomography (CT) has been widely used in clinical, industrial and other applications since its ability to achieve the inner vision of an object without destructing it. With the increased usage of CT in clinics, the potential risk that inducing cancers by the X-ray radiation has been alarmed \cite{ISI:000312590500022}. Therefore, many techniques have been developed to decrease the radiation dose of CT including lowering the x-ray exposure in each tube by decreasing the current and shortening the exposure time of the tubes, and decreasing the number of scanning angles. Lowering the x-ray exposure will result in a noisy sinogram while decreasing the number of scanning angles will make the system ill posed and the reconstructed CT image will suffer from undesirable artifacts. To address these issues, many algorithms were proposed to improve the quality of the reconstructed CT images, which can be generally divided into three categories: (a) sinogram domain processing, (b) iterative algorithm, and (c) image domain post-processing.

Sinogram domain processing techniques first upsample and denoise the sinograms before converting them to CT images. Balda et al. \cite{ISI:000304911300006} introduced a structure adaptive sinogram filter to reduce the noise of  sinograms. Cao et al. \cite{ISI:000336951800030} proposed a dictionary learning based inpainting method to estimate the missing projection data. Lee et al. \cite{ISI:000460685600002} proposed a deep U-network to interpolate the sinogram for sparse-view CT images.

The iterative algorithms reconstruct a CT image by solving the following model:
$${u^*} = \mathop {\arg \min }\limits_{u \in {R^N}} E(Hu,x) + {\lambda _{\rm{1}}}{R_{\rm{1}}}(u){\rm{ + }}{\lambda _2}{R_{\rm{2}}}(Hu),$$
where $E(Hu,x)$  is a data-fidelity term that forces the solution ${u^{\rm{*}}}$  to be consistent with the measured data  $x$,  $H$ is  Radon transform, ${R_{\rm{1}}}(u)$  and ${R_{\rm{2}}}(Hu)$  are two regularization terms that incorporate the prior knowledge of the data in image domain and sinogram domain, respectively, into the reconstructed image and  ${\lambda _{\rm{1}}},{\lambda _2} \ge 0$ are two trade-off parameters. In the literature, many different forms of ${R_{\rm{1}}}(u)$  and  ${R_{\rm{2}}}(Hu)$ are utilized, such as the total variation (TV) \cite{ISI:000258537000022} and its improved versions \cite{ISI:000337176600009}\cite{ISI:000372039000022}\cite{ISI:000323169200006}, nonlocal means (NLM)\cite{ISI:000269734300001}, dictionary learning \cite{ISI:000310148600002}\cite{ISI:000494433300011}, low-rank \cite{ISI:000340237800001} and  ${l_1}$-norm of the wavelet coefficients    \cite{ISI:000287862300014}.

The post-processing techniques improve the quality of CT images by removing the noise and artifacts in the CT images already reconstructed by other methods (such as FBP). In theory, all the methods of removing noise and artifacts of usual optical images can be applied to the CT image post-processing, such as the NLM \cite{ISI:000329182200029}, dictionary learning \cite{ISI:000241537700021}\cite{ISI:000322775300030}, block-matching 3D (BM3D) \cite{ISI:000281481500009}\cite{ISI:000322020600085}, and so on.

Recently, inspired by the development of deep learning  \cite{GoodBengCour16}\cite{ISI:000347595400010}\cite{ISI:000358218600040}\cite{ISI:000355286600030} in computer vision and natural language processing, many deep-learning (DL) based algorithms  have been proposed for CT reconstruction. Most of them were utilized as a post-processing step to remove noise and artifacts of CT images reconstructed by other techniques. For example, in \cite{ISI:000417913600012}, Wang et al. proposed a residual encoder–decoder convolutional neural network (RED-CNN) to remove the artifacts from the FBP reconstruction. In \cite{ISI:000405701500004}, Jin et al. proposed a deep convolutional network (FBPConvNet) that combines FBP, U-net and the residual learning to remove the artifacts while preserving the image structures. In \cite{ISI:000434302700011}, Zhang et al. proposed a DenseNet and deconvolution based network (DD-Net) to improve the quality of the CT images reconstructed by FBP. In \cite{ISI:000494433300020}, Jiang et al. proposed symmetric residual convolutional neural network (SR-CNN) to enhance the sharpness of edges and detailed anatomical structures in under-sampled CT image reconstructed by the ASDPOCS TV algorithm \cite{ISI:000258537000022}. In \cite{ISI:000434302700007}, a framelet-based deep residual network was proposed to denoise the low-dose CT images. Some of other DL based algorithms learn the mapping between  sinogram and CT image space, which directly decodes sinograms into CT images. For instances, in \cite{ISI:000427937900046}, Zhu et al. proposed a deep neural network AUTOMAP to learn a mapping between sensors and 
images. In \cite{ISI:000489784000020}, Li et al. proposed a deep learning network iCT-Net to address difficult CT image reconstruction problems such as view angle truncation, the view angle under-sampling, and interior problems. In \cite{ISI:000434302700005}, Chen et al. proposed a reconstruction network for sparse-data CT by unfolding an iterative reconstruction scheme up to a number of iterations for data-driven training. 

In this paper, we propose a deep network for sinogram denoising and CT image reconstruction simultaneously. Specifically, our network consists of two cascaded blocks, which are linked by a FBP layer. The former block is responsible for denoising and completing the sinograms while the latter is used to removing the noise and artifacts of the CT images. Different from \cite{ISI:000427937900046} and \cite{ISI:000489784000020}, we utilize the FBP layer to decode the sinograms into CT images instead of using the fully connected layer, which reduces the number of parameters of the network and avoids the overfitting problem.

This paper is organized as follows. Section II describes detailed structure of the proposed network and its training method. Section III presents the experimental results. Discussion and conclusion are given in Section IV.

\section{Method}
Let us first introduce the proposed network.
\subsection{Overall network architecture} 
\begin{figure}[!t]
	\centerline{\includegraphics[width=\columnwidth]{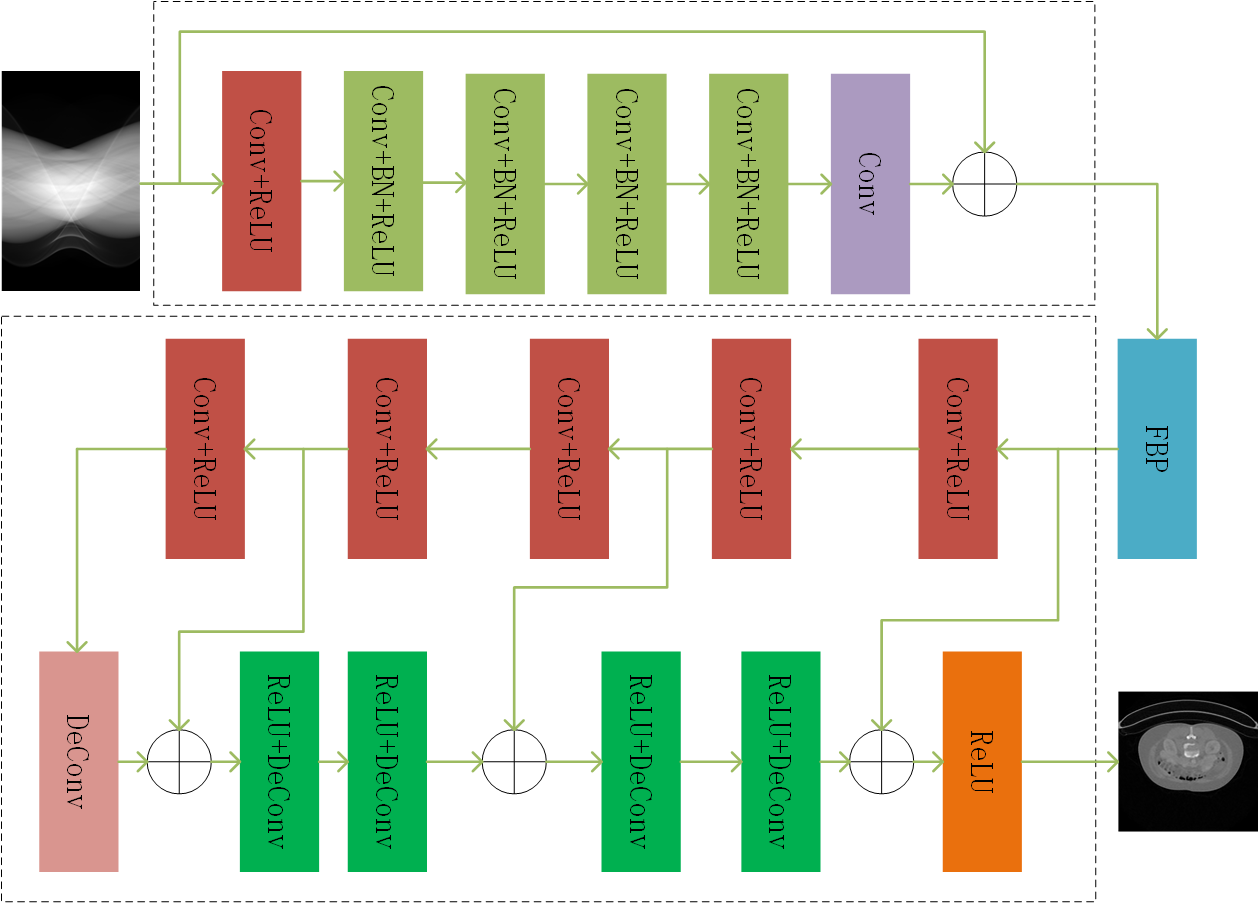}}
	\caption{Overall architecture of our proposed network. It consists of the sinogram domain map $\sigma _{{1}}$ and CT image domain map $\sigma _{{2}}$ which are linked by the FBP layer.}
	\label{fig1}
\end{figure}

Assuming that $u \in {R^{M \times M}}$ is the image of a test object, and $x \in {R^{M \times N}}$ is its corresponding sinogram. The relationship between them can be modeled as
$$x = Pu,$$
where $M$ and $N$ are the number of detectors and projection angles, respectively, and $P$ is the measurement process involving with a Radon transform and some noisy factors. Our goal is to use the deep learning (DL) techniques to learn a map \[\sigma :{R^{M \times N}} \to {R^{M \times M}}\] such that $\sigma (x)$ approximates $u$ for all training data pairs $\{ (u,x)\} $. The pipeline of our proposed network has three cascade steps: The first step is to denoise and complete the sinograms followed by the second step to convert the processed sinograms to CT images and the third step to remove the noise and artifacts resided in the CT images. This general process can be expressed as ${\sigma _{{2}}}\left( {F({\sigma _{{1}}}(x))} \right)$, where 
$${\sigma _{{1}}}:{R^{M \times N}} \to {R^{M \times N}}$$ 
is the sinogram domain map function (The numbers of the angles of ${\sigma _{{1}}}(x)$ are still $N$ since we interpolate $x$ along the angle direction before inputting it to ${\sigma _{{1}}}$), 
$${\sigma _{{2}}}:{R^{M \times M}} \to {R^{M \times M}}$$ 
is the image domain map function and 
$$F:{R^{M \times N}} \to {R^{M \times M}}$$
is the FBP transform layer that decodes sinograms into CT images. When designing layers that map elements from sinograms space (${R^{M \times N}}$) to CT image space (${R^{M \times M}}$), fully connected layers are usually utilized \cite{ISI:000427937900046} \cite{ISI:000489784000020}, which increases the size of the parameters and makes the network prone to be overfit. Therefore, in our network we use the FBP layer to replace the fully connected layers. Thus, learning one single map \[\sigma :{R^{M \times N}} \to {R^{M \times M}}\]
 is converted into learning two maps 
 $${\sigma _{{1}}}:{R^{M \times N}} \to {R^{M \times N}}$$ 
 and 
 $${\sigma _{{2}}}:{R^{M \times M}} \to {R^{M \times M}},$$
  which is easier since the latter two functions ${\sigma _{{1}}}$ and ${\sigma _{{2}}}$ map elements between the same spaces. The overall architecture of our network can been seen in Figure \ref{fig1}.

\subsection{Network architecture for preprocessing sinograms}
We construct our sinogram domain network ${\sigma _{{1}}}$ based on the residual DnCNN \cite{ISI:000401297400005}, which was originally developed to remove blind Gaussian noise. According to \cite{ISI:000400012300083}, residual learning in DnCNN makes the residual mapping much easier to be optimized since it is more like an identity mapping. In our paper, the architecture of network ${\sigma _{{1}}}$ also consists of three types of layers. For the first layer, convolution unit with filters of size ${{3}} \times {{3}} \times {{1}} \times {{64}}$ is used to generate 64 features, and activation unit ReLU $(f(x) = \text{max}(x,0))$ is then used for nonlinearity. For Layers ${{2}} \sim {{5}}$, convolution units, batch normalized (BN) units and activation units ReLU are used, where the filter sizes of convolutions are $3 \times 3 \times {{64}} \times 64$. For Layer 6, a single convolution unit with filters of size $3 \times 3 \times {{64}} \times {{1}}$ is used to reconstruct the sinograms. At last, a shortcut is utilized to connect the input and output.

\subsection{Network architecture for post-processing CT images}
In the literature, large amounts of deep networks were proposed to post-process the CT images reconstructed by other methods. In our network, the Red-CNN \cite{ISI:000417913600012} is adopted to construct our CT image domain network ${\sigma _{{2}}}$. The detailed structures about the network are as follows. For Layers ${{8}} \sim {{12}}$ (Layer 7 is the FBP), convolution units are used to generate feature maps and activation units ReLU are used for nonlinearity, where the filter size of Layer 8 is ${{5}} \times {{5}} \times {{1}} \times {{96}}$ and ${{5}} \times {{5}} \times {{96}} \times {{96}}$ of Layers ${{9}} \sim {{12}}$. For Layers ${{13}} \sim {{18}}$, deconvolution units are used to decode features and activation units ReLU are used for nonlinearity, where the filter size of Layers ${{13}} \sim {{16}}$ is ${{5}} \times {{5}} \times {{1}} \times {{96}}$ and ${{5}} \times {{5}} \times {{96}} \times {{1}}$ of Layer ${{17}}$, and the strides of the deconvolutions are all 1. Also, there are 3 shortcuts that connect the FBP layer and the deconvolution units of Layer 17, the ReLU units of Layer 9 and the deconvolution units of Layer 15, and the ReLU units of Layer 11 and the deconvolution units of Layer 13, respectively.  

\subsection{Loss function and training}

The loss function of our whole network is composed of two terms, 
\begin{equation}\label{E1}
L(x,\theta ) = {L_{{\sigma _1}}} + {L_{{\sigma _{{2}}}}}
\end{equation}
where
$${L_{{\sigma _1}}}= \frac{\lambda }{N}\sum\nolimits_{i = 1}^N {{{\left\| {{y_i} - {\sigma _1}({x_i};{\theta _1})} \right\|}^2}},$$
$${L_{{\sigma _{{2}}}}}=\frac{{(1 - \lambda )}}{N}\sum\nolimits_{i = 1}^N {{{\left\| {{u_i} - {\sigma _{{2}}}\left( {F({\sigma _{{1}}}({x_i};{\theta _1}));{\theta _2}} \right)} \right\|}^2}}, $$
 $\theta {{ = }}\left\{ {{\theta _1},{\theta _2}} \right\}$ are the parameters to be learned, ${{\{ }}{x_i},{y_i},{u_i}{{\} }}$ is an set of associated data, ${x_i}$ represents the input sinogram, ${y_i}$ is a sinogram label and ${u_i}$ is a CT image label, ${\sigma _1}({x_i})$ is the output of the sinogram domain network and ${\sigma _{{2}}}\left( {F({\sigma _{{1}}}({x_i}))} \right)$ is the output of the whole network, and $\lambda  \geqslant {{0}}$ is a balance parameter.
Our network is an end-to-end system  mapping sinograms to CT images. Once the architecture of the network is configured, its parameters can be learned by optimizing the loss function (\ref{E1}) using the backpropagation algorithm (BP) \cite{INSPEC:3589863}. In this study, the loss function is optimized by the Adam algorithm \cite{kingma2014adam}, where the learning rate was set as ${{1}}{{{0}}^{ - 3}}$.

\subsection{Gradients and backpropagation}
In our experiments, we use the software Tensorflow to train the network and compute the gradients of the loss function with respect to its parameters. During the training process, two main gradients need to be calculated, $\frac{{\partial L}}{{\partial {\theta _{{1}}}}}$ and $\frac{{\partial L}}{{\partial {\theta _{{2}}}}}$, where $\frac{{\partial L}}{{\partial {\theta _{{2}}}}}$ can be computed by Tensorflow automatically while $\frac{{\partial L}}{{\partial {\theta _{{1}}}}}$ needs more efforts. By the chain rules, we have 
\begin{equation*}
\begin{aligned}
&\frac{{\partial L}}{{\partial {\theta _{\rm{1}}}}} = \frac{{\partial {L_{{\sigma _{\rm{1}}}}}}}{{\partial {\theta _{\rm{1}}}}} + \frac{{\partial {L_{{\sigma _{\rm{2}}}}}}}{{\partial {\theta _{\rm{1}}}}}\\
= &\frac{{\partial {L_{{\sigma _{\rm{1}}}}}}}{{\partial {\theta _{\rm{1}}}}}\; - \frac{{2(1 - \lambda )}}{N}\sum\nolimits_{i = 1}^N {\frac{{\partial {L_{{\sigma _{\rm{2}}}}}}}{{\partial {\sigma _{\rm{2}}}}}\frac{{\partial {\sigma _{\rm{2}}}\left( {F({\sigma _{\rm{1}}}({x_i};{\theta _1}));{\theta _2}} \right)}}{{\partial {\theta _2}}}} \\
= &\frac{{\partial {L_{{\sigma _{\rm{1}}}}}}}{{\partial {\theta _{\rm{1}}}}}\\ 
- &\frac{{2(1 - \lambda )}}{N}\sum\nolimits_{i = 1}^N {\frac{{\partial {L_{{\sigma _{\rm{2}}}}}}}{{\partial {\sigma _{\rm{2}}}}}\frac{{\partial {\sigma _{\rm{2}}}(F( \bullet );{\theta _2})}}{{\partial F}}\frac{{\partial F({\sigma _{\rm{1}}}( \bullet ))}}{{\partial {\sigma _{\rm{1}}}}}\frac{{\partial {\sigma _{\rm{1}}}({x_i};{\theta _1})}}{{\partial {\theta _1}}}} 
\end{aligned}
\end{equation*}

where \[\frac{{\partial {L_{{\sigma _{{1}}}}}}}{{\partial {\theta _{{1}}}}}{{ = }} - \frac{{{{2}}\lambda }}{N}\sum\nolimits_{i = 1}^N {({y_i} - {\sigma _1}({x_i};{\theta _1}))\frac{{\partial {\sigma _1}({x_i};{\theta _1})}}{{\partial {\theta _1}}}},\]
\[\frac{{\partial {L_{{\sigma _{{2}}}}}}}{{\partial {\sigma _{{2}}}}}{{ = }}{u_i} - {\sigma _{{2}}}\left( {F({\sigma _{{1}}}({x_i};{\theta _1}));{\theta _2}} \right),\]
and $F({\sigma _{{1}}}({x_i};{\theta _1}))$ is the output of the FBP layer that decodes the sinograms into CT images. 

The gradients ${\frac{{\partial {\sigma _{\rm{2}}}(F( \bullet );{\theta _2})}}{{\partial F}}}$ and ${\frac{{\partial {\sigma _{\rm{1}}}({x_i};{\theta _1})}}{{\partial {\theta _1}}}}$ can be calculated by Tensorflow automatically.
For the FBP layer,  using “@tf.function” in Tensorflow and “iradon” function in “scikit-image” package can constitute it. However, when automatically computing the gradient $\frac{{\partial F({\sigma _{{1}}}(\bullet))}}{{\partial {\sigma _{{1}}}}}$ of this version of FBP layer by Tensorflow, an error will be raised. To solve this issue, we implement the FBP layer by using the sparse matrix multiplication. 

Let $s \in {R^{1 \times MN}}$ be the vectorization $x_v$ of a sinogram $x \in {R^{M \times N}}$, i.e. 
$$s =x_v= \text{reshape}(x,[1,MN]),$$ 
then there exists a real sparse matrix ${B} \in {R^{MM \times MN}}$ such that the matrix multiplication ${B}s$ equals to the vectorization of the backprojection of $x$.  Since the FBP method reconstructs  CT images by convoluting with sinograms followed by a backprojection, the output $F(x)$ of our FBP layer for input $x$ can be rewritten as
$$F(x) = \text{reshape}({B}\tilde x_v,[M,M]),$$
where 
$$\tilde x_v=\text{reshape}(x \otimes h,[1,MN]),$$
$h \in {R^M}$ is the “ramp filter”, i.e. $\hat h(w) = \left| w \right|$,  $\hat h$ is the Fourier transform of $h$, and $x \otimes h$ represents that each columns of $x$ convolutes with $h$ circularly. 
After constructing the FBP layer by using the sparse matrix $B$ and filter $h$, we can compute the gradient $\frac{{\partial F(x)}}{{\partial x}}$.
\begin{align*}
\frac{{\partial F(x)}}{{\partial {x_{i,j}}}} =& \frac{{\partial \text{reshape}({B}\tilde x_v,[M,M])}}{{\partial {x_{i,j}}}} \hfill \\
= &\text{reshape}({B}\tilde E^{i,j},[M,M]), \hfill 
\end{align*}
where $$\tilde E^{i,j}=\text{reshape}({E^{i,j}} \otimes h,[1,MN]),$$
${x_{i,j}}$ is the $(i,j)$ entry of $x$ and ${E^{i,j}}\in R^{M\times N}$ is a matrix with its $(i,j)$ entry being 1 and others 0.

\section{Experimental results}
We now present some simulation results.
\subsection{Data preparation}
\begin{enumerate}
	\item Train dataset. A Clinical Proteomic Tumor Analysis Consortium Lung Adenocarcinoma (CPTAC-LUAD) \cite{data1} dataset was downloaded from The Cancer Imaging Archive (TCIA). We randomly chose 500 CT images from CPTAC-LUAD, extracted their central patch of size ${{256}} \times {{256}}$ and stretched the value in the interval $[0,255]$ linearly. The 500 extracted patches were then used as the CT image labels ${u_i}$ and the sinogram image labels ${y_i}$ were generated by 
	$${y_i} = H({u_i}),$$
	 where $H$ is Radon transform. The input sinograms ${x_i}$ were generated by adding noise to ${y_i}$ via the following equations \cite{ISI:000434302700019}:
	 \begin{equation}\label{E2}
	 \begin{aligned}
	 {y_p} &= \frac{1}{b}\text{Poisson}(b\exp ( - \frac{{{y_i}}}{{\max ({y_i})}}))\\
	 {y_g} &= \text{Gaussian}(\operatorname{var} )\\
	 {z_i} &= \text{min} ({y_p} + {y_g},1)\\
	 {x_i} &= \text{max} ({y_i})*\log ( - {z_i})
	 \end{aligned} 
	 \end{equation}
	 
	where we set $b = {10^{{7}}}$ and $\operatorname{var}  = 0.002$ in our experiments.
	\item Test dataset. A Pancreas-CT \cite{data2} dataset was downloaded from TCIA. The Pancreas-CT contains 82 abdominal contrast enhanced 3D CT scans from 53 male and 27 female subjects. Since the CT images in Pancreas-CT are of size ${{512}} \times {{512}}$, we downsampled them by factor 2 and randomly chose 500 of them as the test data. The input sinograms of the test set were generated in the way as generating the training sinograms ${x_i}$ via equation (\ref{E2}).
\end{enumerate}
\subsection{Reconstructed results for 180 angles}
\begin{figure}[!t]
	\centering{
	\subfloat[Original]{\includegraphics[width=0.32\columnwidth]{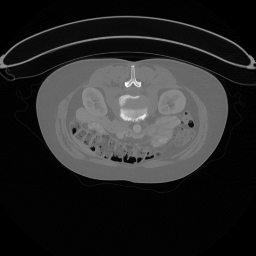}}
	\subfloat[FBP]{\includegraphics[width=0.32\columnwidth]{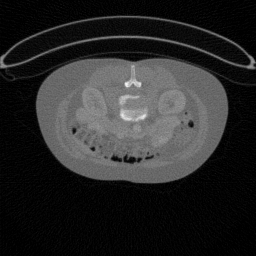}}
    \subfloat[Red-CNN]{\includegraphics[width=0.32\columnwidth]{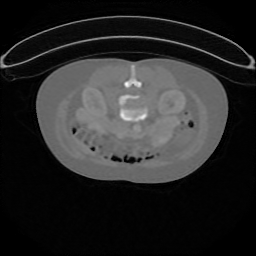}}\\
	\subfloat[DD-Net]{\includegraphics[width=0.32\columnwidth]{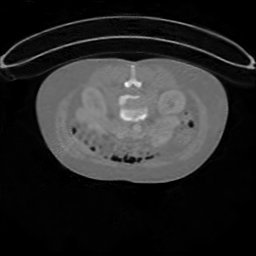}}
	\subfloat[FBP-Conv]{\includegraphics[width=0.32\columnwidth]{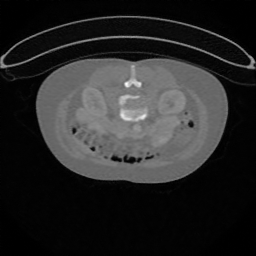}}
	\subfloat[Ours]{\includegraphics[width=0.32\columnwidth]{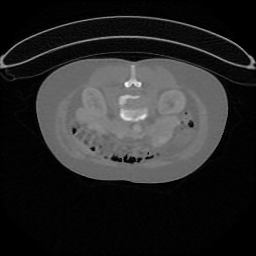}}
}
	\caption{The reconstructed results of the compared methods for 180 angles.}
	\label{FigB1}
\end{figure}

\begin{figure}[!t]
	\begin{center}
	\subfloat[Red-CNN]{\includegraphics[width=0.32\columnwidth]{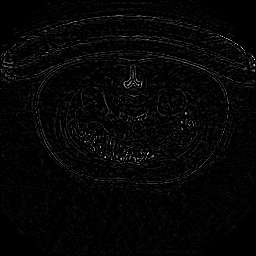}}
	\subfloat[DD-Net]{\includegraphics[width=0.32\columnwidth]{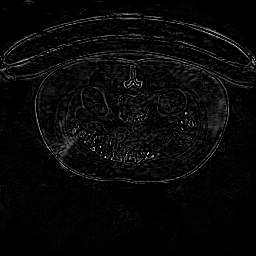}}\\
	\subfloat[FBP-Conv]{\includegraphics[width=0.32\columnwidth]{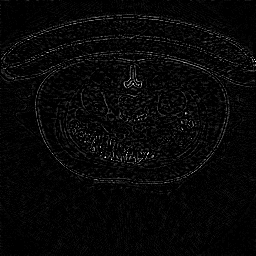}}
	\subfloat[Ours]{\includegraphics[width=0.32\columnwidth]{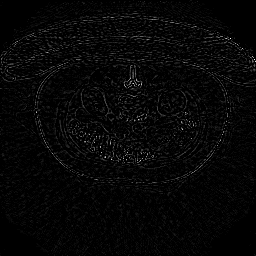}}
\end{center}
	\caption{The absolute difference images  between the reconstructed and the original for 180 angles.}
	\label{FigB2}
\end{figure}

In this subsection, we demonstrate that our network can be trained to reconstruct CT images from the sparse-viewed angle sinograms. To this end, we sampled the sinograms at 180 angles ([0:1:180]) that are uniformly spaced in the interval $[0,\pi ]$ to get the simulated data ${x_i}$ and ${y_i}$. Therefore, in this set of experiments, the number $N$ of angles of the input sinograms ${x_i}$ and sinogram image labels ${y_i}$ are both 180. For comparison, referenced CT images are reconstructed by state of the art deep learning methods, Red-CNN \cite{ISI:000417913600012}, DD-Net \cite{ISI:000434302700011} and FBP-Conv \cite{ISI:000405701500004}. For these comparted methods, we use the standard FBP algorithm with the “Ram-Lak” filter to reconstruct the CT images from the sinograms ${x_i}$ as their inputs. We set $\lambda=0.1$ in our loss function for this set of experiments.
In Figure \ref{FigB1}, the compared results reconstructed from the test data are shown, from which we can observe that the reconstructed result by DD-Net still has some artifacts while those reconstructed by the other methods have similar visual effect. In Figure \ref{FigB2}, we present the absolute difference images between the reconstructed and the original. We can see that the reconstructed image by our method lost the least details compared to those by the other methods.

To evaluate the performance of these networks objectively, PSNR and SSIM are used to measure the similarity of the reconstructed images and the original. In Table \ref{T1}, the average values of PSNR and SSIM of the results reconstructed from the test dataset by the five methods (including FBP) are listed, from which we can observe that our network gets the highest PSNR and SSIM in average.
\begin{table}[htbp]
	\centering
	\caption{The averaged PSNR and SSIM of the compared methods for 180 angles.}
	\centering
	\begin{tabular}{c|c|c}
		\hline
		& PSNR  & SSIM   \\ \hline
		FBP	     &32.21 	&0.788 \\
		Red-CNN	 &36.08 	&0.929  \\
		DD-Net	 &34.24 	&0.889  \\
		FBP-Conv &35.74 	&0.928  \\
		Ours     &\textbf{36.33 }    &\textbf{0.933}   \\
		\hline
	\end{tabular}%
	\label{T1}
\end{table}%
\subsection{Reconstructed results for 90 angles}
\begin{figure}[!t]
	\centering{
		\subfloat[Original]{\includegraphics[width=0.32\columnwidth]{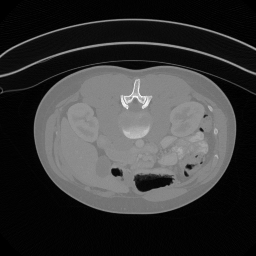}}
		\subfloat[FBP]{\includegraphics[width=0.32\columnwidth]{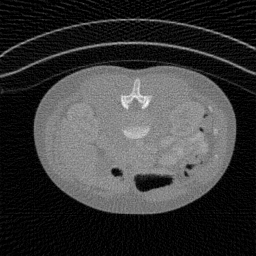}}
		\subfloat[Red-CNN]{\includegraphics[width=0.32\columnwidth]{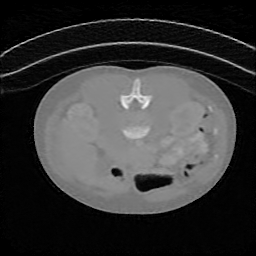}}\\
		\subfloat[DD-Net]{\includegraphics[width=0.32\columnwidth]{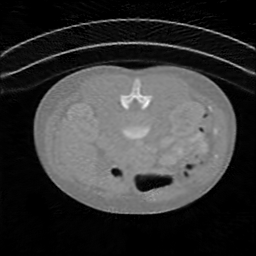}}
		\subfloat[FBP-Conv]{\includegraphics[width=0.32\columnwidth]{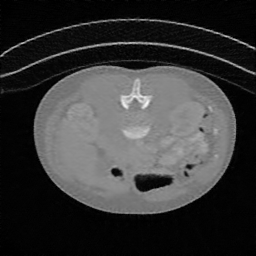}}
		\subfloat[Ours]{\includegraphics[width=0.32\columnwidth]{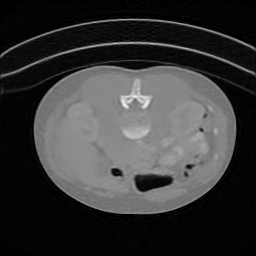}}
	}
	\caption{The reconstructed results of the compared methods for 90 angles.}
	\label{FC1}
\end{figure}

\begin{figure}[!t]
	\centering{
		\subfloat[Red-CNN]{\includegraphics[width=0.32\columnwidth]{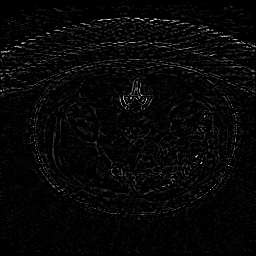}}
		\subfloat[DD-Net]{\includegraphics[width=0.32\columnwidth]{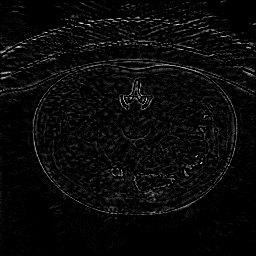}}\\
		\subfloat[FBP-Conv]{\includegraphics[width=0.32\columnwidth]{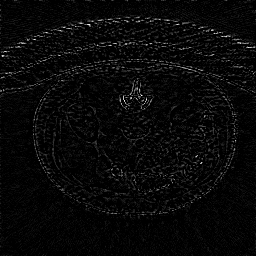}}
		\subfloat[Ours]{\includegraphics[width=0.32\columnwidth]{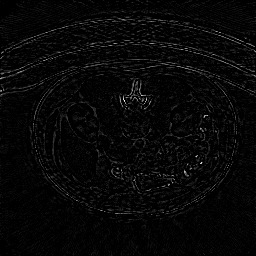}}}
	\caption{The absolute difference images  between the reconstructed and the original for 90 angles.}
	\label{FC2}
\end{figure}

In this subsection, sparser sinogram data are used to examine the ability of our network to reconstruct CT images. We first sampled the sinograms at 90 angles ([0:2:180]) that are uniformly spaced in the interval $[0,\pi ]$, then add noise to the samples via equation (\ref{E2}) and at last interpolate them along the angle direction to 180 angles to get the input sinograms ${x_i}$. The sinogram image labels ${y_i}$ were obtained by sampling the sinograms at 180 angles ([0:1:180]). Thus, the actual number of angles of the input sinograms ${x_i}$ is 90 while the one of the labels ${y_i}$ is 180. We set $\lambda {{ = 0}}{{.5}}$ in our loss function for this set of experiments. We also compared our results to those of Red-CNN \cite{ISI:000417913600012}, DD-Net \cite{ISI:000434302700011} and FBP-Conv \cite{ISI:000405701500004}.
Figure \ref{FC1} shows the results of the compared methods reconstructed from the test dataset. We can see that the results of DD-Net and FBP-Conv have some noise and artifacts while those of Red-CNN and ours have the best visual effect. Similarly, we display the absolute difference images between the reconstructed results and the original in Figure \ref{FC2}, from which we can observe that the result of our network preserve more details. 

Quantitative analysis for the reconstructed results of the entire test data using these methods has also been carried out. The average PSNR and SSIM of the results are listed in Table \ref{T2}. We can observe that our network clearly outperforms the other methods and has the highest PSNR and SSIM in average.
\begin{table}[htbp]
	\centering
	\caption{The averaged PSNR and SSIM of the compared methods for 90 angles.}
	\centering
	\begin{tabular}{c|c|c}
		\hline
		& PSNR  & SSIM   \\ \hline
		FBP	     &28.33 	&0.593 \\
		Red-CNN	 &33.58 	&0.896  \\
		DD-Net	 &32.05 	&0.833  \\
		FBP-Conv &33.84 	&0.898  \\
		Ours     &\textbf{34.39 }    &\textbf{0.918 }  \\
		\hline
	\end{tabular}%
	\label{T2}
\end{table}%
\subsection{Effect test of parameter $\lambda$}
In this subsection, we test the effect of the parameter $\lambda $ in our loss function. First, we test its effect on the sinograms of 180 angles. We set $\lambda =0.1,~0.5,~0.9$ to train the network with the training data of 180 angles. The average PSNR and SSIM of the sinograms and CT images reconstructed from the test dataset are listed in Table \ref{T3}. From Table \ref{T3}, we can see that the network with $\lambda {{ = 0.1}}$ can reconstruct  the best CT images but its subnetwork ${\sigma _{{1}}}$ has least denoising effect. Conversely, the subnetwork ${\sigma _{{1}}}$  with $\lambda {{ = 0}}{{.9}}$ has good denoising effect but the average PSNR of its results is lower than that of $\lambda = 0.1$.

Next, we train our network using the training data of 90 angles with $\lambda =0.1,~0.5,~0.9$. The average PSNR and SSIM of the reconstructed sinograms and CT images with different $\lambda$ are listed in Table \ref{T4}. We can observe that the subnetwork ${\sigma _{{1}}}$ with large $\lambda$ can output sinograms of better quality. But the CT images reconstructed by the whole network with $\lambda=0.5$ have the highest average PSNR, which is different from that using sinograms of 180 angles. This may be because that when we train the network using sinograms of 90 angles, the corresponding labels we used are of 180 angles, which may reconstruct better sinograms and  the better sinograms output by the network with $\lambda=0.5$ also have a positive effect on the final CT images reconstruction. 
\begin{table}[htbp]
	\centering
	\caption{The effect test of $\lambda$ on sinograms of 180 angles.}
	\centering
	\begin{tabular}{c|cc|cc}
		\hline
		& \multicolumn{2}{c}{Sinograms}\vline &\multicolumn{2}{c}{CT images}\\
		\hline
		& PSNR  & SSIM  & PSNR  & SSIM \\ \hline
		$\lambda=0.1$     &51.43 	&0.993 &36.33&0.933\\
		$\lambda=0.5$	 &56.00  	&0.997 &36.13&0.931\\
		$\lambda=0.9$	 &56.12 	&0.998 &36.28&0.933\\
		\hline
	\end{tabular}%
	\label{T3}
\end{table}%

\begin{table}[htbp]
	\centering
	\caption{The effect test of $\lambda$ on sinograms of 90 angles.}
	\centering
	\begin{tabular}{c|cc|cc}
		\hline
		& \multicolumn{2}{c}{Sinograms}\vline &\multicolumn{2}{c}{CT images}\\
		\hline
		& PSNR  & SSIM  & PSNR  & SSIM \\ \hline
		$\lambda=0.1$     &52.96 	&0.996  &34.15&0.918\\
		$\lambda=0.5$	 &53.27 	&0.996  &34.39&0.918\\
		$\lambda=0.9$	 &56.12 	&0.996  &34.32&0.917\\
		\hline
	\end{tabular}%
	\label{T4}
\end{table}%

\section{Conclusion}
In this paper, we  proposed an end-to-end deep network for CT image reconstruction, which inputs the sinograms and outputs the reconstructed CT images. The network consists of two blocks, which are linked by an FBP layer.
The former block pre-processes the sinograms such as denoising and upsampling, the latter block post-processes the CT images such as denoising and removing the artifacts  and the FBP layer decodes the sinograms into CT images. By using the sparse matrix multiplication, the problem that computes the gradients of the FBP layer with respect to the parameters of the first block was addressed. Experimental results demonstrated that our method outperforms state of the art deep learning method in CT reconstruction. One reason why the performance of our network is better than others is that we train our network using extra information, i.e. the sinogram image labels.

%
\bibliographystyle{IEEEtran}
\bibliography{merged}
\end{document}